# EMISSION LINE PROFILES FROM SELF-GRAVITATING THIN DISKS

V. Karas,[1,3] A. Lanza,[1] and D. Vokrouhlický[2,3]

To appear in The Astrophysical Journal (enlarged version)

[1]Scuola Internazionale Superiore di Studi Avanzati, Via Beirut 4, I-34014 Trieste, Italy

[2]Observatoire de la Côte d'Azur, dept. CERGA, Av. Nicolas Copernic, F-06130 Grasse, France

[3]On leave from the Astronomical Institute, Charles University, Prague
Internet: karas@tsmi19.sissa.it; lanza@tsmi19.sissa.it; david@ocar01.obs-azur.fr

## ABSTRACT

We have constructed general relativistic models of a thin disk around a rotating black hole and computed profiles of a spectral line, emitted in the inner region of the disk. In our models we have taken into account also the self-gravity of the disk. The aim of this work is to study gravitational effects on the line profiles in connection with the X-ray features observed in spectra of active galactic nuclei. In some cases, the calculated profiles are clearly affected by the disk gravity but relativistic dragging effects are found to be negligible.

*Subject headings:* accretion disks — galaxies: active — $\gamma$-ray sources — black hole physics

## 1. INTRODUCTION

Accretion disks play an important role in the theory of various astronomical objects. Observations confirm the presence of accretion disks in close binary stars (Pringle & Wade 1985; Shore, Livio, & van den Heuvel 1994) and they are also believed to be the central source of energy in active galactic nuclei (AGN). Despite the great effort of both observers and theorists, the current evidence for accretion disks in AGN remains only indirect (Frank, King, & Raine 1992; Osterbrock 1993). Theoretical models do not explain observational evidence in a unique way. Still it is highly desirable to have at least order of magnitude estimates of various effects for comparison with the observational data. In this paper, the effects of the gravity of the system on the spectral line emitted in the inner region of a disk are studied. In particular, the effects of the self-gravity of the disk on the motion of the photons emitted from the surface of the disk are discussed within the framework of general relativity. Very little is known about the mass of the disks in AGN, but some estimates suggest that their self-gravity could be important (Wiita 1982; Abramowicz et al. 1984; Shlosman & Begelman 1987). Even weakly self-gravitating disks may affect the line profiles significantly if the source is close to the black hole horizon or if the inclination of the observer is large. Self-gravitating axially symmetric black hole–disk configurations may also be relevant for the theory of possible stellar-mass neutron tori around neutron stars which have been proposed as a source of $\gamma$-ray bursts (Jaroszyński 1993; Witt et al. 1994). The mass $m$ of neutron tori (in units of the total mass of the system) is even more uncertain than that of the disks in AGN.

According to the current 'standard model' of AGN, the inner parts of the disk are geometrically thin and optically thick. In the outer region, matter forms a torus which can obscure the radiation from the centre. We have assumed that the lines are formed within the inner region. We did not consider the effects of an obscuration in our analysis since they introduce another free parameter, the shape of the torus, into the problem (cf. Kojima & Fukue 1992; Bao & Stuchlík 1992) and this complication will be discussed separately elsewhere. In order to distinguish the effects of light bending from rather complicated microphysics of the accretion process, we have adopted a very simplified and well understood model of generation of radiation from a thin disk—the standard disk model (Novikov & Thorne 1973; Page & Thorne 1974). Within this framework the emitted radiation is determined by rather general conservation laws, but now with our numerically constructed spacetime. The restriction on the disk thickness is not crucial for our method, but it allows a comparison with related works on the line profiles and reveals features which can be attributed to the disk self-gravity. It would not be too difficult to accommodate, for example, the model of ion-supported tori (Rees et al. 1982) if all the parameters that determine the shape of the torus are predetermined. In other words, we have solved Einstein's equations for an axially symmetric spacetime describing a thin, stationary, Keplerian disk around a rotating black hole and we have computed distortions of spectral line profiles originating at the disk and observed at a large distance. Our approach can be generalized to construct geometrically thick tori and to calculate their spectra. In addition, the mass of the disk is not restricted to negligible values *a priori*, apart from the fact that our thin Keplerian disks become unstable against adiabatic convective perturbations (Seguin 1975) for $m \gtrsim 0.1$.

In some situations (e.g. when the inclination angle of the observer is large) the disk self-gravity affects the photon motion significantly. However, it is not apparent whether significant modifications of spectra are to be expected in comparison with models in which the self-gravity of the disk is negligible. In particular the role of higher order images can be more important. Our present contribution supplements several works of other authors who studied problems of the line formation in accretion disks but ignored their self-gravity: Gerbal & Pelat (1981) study line profiles emitted by a Keplerian ring orbiting a Schwarzschild black hole. Laor (1991) generalizes this discussion to the case of a ring around a rotating (Kerr) black hole. Laor & Netzer (1989) and Kojima (1991) examine radiation from a thin disk. Kojima & Fukue (1992) consider tori with different shapes and angular momentum distribution but they ignore general relativistic effects. Bao & Stuchlík (1992) study effects of the disk self-eclipses when the observer inclination is large enough within the framework of the Schwarzschild metric. They mention the presence of a central peak in the line profiles, whose detection is however beyond the current observational capabilities. Bao, Hadrava & Østgaard (1994) study the role of the first order photons and they conclude that their contribution should not be ignored, of course provided that these photons are not reprocessed by the disk material as it is usually assumed. Chen, Halpern, & Filippenko (1989), Chen & Halpern (1989a, b), and Halpern (1990) attempt to fit computed line profiles from a Keplerian disk around Schwarzschild black hole to the actually observed lines from several objects. Besides the radiative flux, general relativistic effects modify polarization properties of the radiation (Connors, Piran & Stark 1980). Chen & Eardley (1991) investigate polarization features of spectral lines emitted from a thin disk. In this model polarization is due to electron scattering in the disk corona, again within the Schwarzschild geometry. Polarization studies may provide further information about the gravitational field of the source, however, at present the structure of the disk corona in AGN is not well understood. A very interesting review of observational data relevant for the lines produced at intermediate distances from the centre has been recently given by Eracleous & Halpern (1994; This study considers spectral lines in the optical region rather than the X-ray region which we consider here).

At the present there is intensive interest in time-scales for both continuum and line variability of AGN. Mushotzky, Done, & Pounds (1993) give a review of recent observational evidence for variability in the X-ray band (0.1–100 keV) which appears particularly relevant for our discussion. Adopting the accretion disk model of AGN, feature-less X-ray continuum variability on short time-scales, $10^2$–$10^3$ sec, indicates that most of the radiation comes from the innermost regions located at $\lesssim 30$ gravitational radii from the black hole. Several spectral features were detected in a few keV range, in particular fluorescent iron lines which have been recently discussed by a number of authors; see Treves, Perola, & Stella (1991) and references cited therein. It appears that 6.7 keV and 6.9 keV iron emission lines are formed in the inner region while $\approx$ 6–7 keV fluorescence line dominates farther outside (Kallman & White 1989). This scheme has not been fully developed yet and, taking into account the low resolution of currently available data ($E/\Delta E \approx 10$), it is difficult to determine the region where the lines are formed. Information about the spectral line can be obtained from the centroid energy, the full-width-at-half-maximum, the equivalent width, the number of peaks and the shape of wings of the line (Treves et al. 1991). In addition, time-lags between variations of the continuum and the corresponding response of the line intensity give constraints on the region where the lines are formed (Stella 1990). Several authors (Fabian et al. 1989; Stella & Campana 1991; Matt, Perola, & Piro 1991; Matt et al. 1992) have studied iron line profiles in AGN. It is evident that further data on the line variability and better resolution of the line profiles are crucial in order to eliminate ambiguities in the interpretation of various spectral features.

This paper is organized as follows: in Section 2. we describe the method of calculation and the physical assumptions of our model. In this section we discuss also the contruction of the astrophysically relevant models of general relativistic disks around a rotating black hole. Then the computation of a typical line profile is described. The results of such calculations and the conclusions are summarized in Section 3. In Appendix A we give additional technical details relevant to the integration of the photon trajectories.

## 2. CALCULATION OF THE SPECTRAL LINE PROFILES

The aim of this study is to calculate synthetic profiles of a spectral line, the source of which is a self-gravitating thin disk around a rotating black hole. In order to do this, one has to (i) specify the spacetime metric; (ii) transfer the radiative intensity along the photon rays from the source to the observer (which is assumed to be in an asymptotically flat region); and (iii) specify the details of the source emissivity in its local rest frame. The main body of our study lies in the careful treatment of the first task. We have employed exact solutions of Einstein's equations representing rotating black holes with a self-gravitating disk around it. These solutions are given numerically. The self-gravity of the disk affects the propagation of light and the corresponding gravitational redshift. The local structure of the disk and its emissivity are also different from the case when the disk gravity is ignored. (Naturally, the contributions to the spacetime structure from the black hole and from the disk cannot be strictly separated within the framework of exact relativistic solutions.) The prescription for the radiative transfer along null geodesics is straightforward because our spacetime is vacuum outside the disk and we adopted the approximation of geometrical optics. However, in this case, the procedure of solving the geodesic equation is rather involved technically, because the metric itself is given as a numerical solution. The detailed and realistic description of the disk local emissivity requires a consistent solution of the disk vertical structure and of its microphysics, however, this is beyond the scope of the present paper. Our aim is to describe relativistic effects on the light trajectories. To overcome the problem of unknown details of the disk structure we have adopted a simplified description of the disk.

### 2.1. Construction of the metric

The numerical procedure used to construct the metric employs a multigrid algorithm (Brandt & Lanza 1988, Lanza 1992a, b). It is based on the idea of iterating successive approximate solutions in grids with different mesh-sizes until convergence is reached. This paragraph summarizes the important features of the method. We assumed that the spacetime is stationary and axially symmetric. The metric in usual notation (Misner, Thorne & Wheeler 1973) takes the form

$$ds^2 = -e^{2\tilde{\nu}}\,dt^2 + e^{2\psi}\,(d\phi - \omega dt)^2 + e^{2\mu}\,\left(dr^2 + r^2 d\theta^2\right)\,, \tag{1}$$

where the metric functions $\tilde{\nu}$, $\psi$, $\omega$, and $\mu$ depend only on the polar spherical coordinates $r$ and $\theta$. The convenient radial coordinate $r$ was originally introduced by Bardeen (1973b). The unknown metric functions are determined by the solution of Einstein's equations in the form

$$G_{\alpha\beta} = 8\pi T_{\alpha\beta}\,, \tag{2}$$

where $T_{\alpha\beta} = \sigma u_\alpha u_\beta$ is the stress–energy tensor of the disk material, and $\sigma(r)$ is the surface energy density defined as

$$\sigma(r) = 2\int_0^{\pi/2} \epsilon\, e^{2\mu}\, r\, d\theta \tag{3}$$

($\epsilon$ is the total energy density). To these equations, one should add the appropriate boundary conditions. In our case, we are interested in solutions describing a rotating black hole surrounded by an equatorial disk ($\theta = \pi/2$, $r_{\rm in} \leq r \leq r_{\rm out}$). Therefore our domain extends from a regular horizon to asymptotically flat region. The mass distribution of the disk is specified by jump conditions (Bardeen & Wagoner 1971) for the energy-momentum tensor in the $\theta = \pi/2$ plane, which also provide boundary conditions in the plane. The initial guess for numerical solution is taken to be the Kerr spacetime. At convergence, the code produces a solution which corresponds to a perfect fluid, pressure-less disk around a rotating black hole. This solution can be characterized by the total mass $M$ and the total angular momentum $J$, defined as Komar's integrals in the resulting metric

$$M = 8\pi \int (T^t_t - \tfrac{1}{2} T^\alpha_\alpha)\, dV \ , \quad J = 8\pi \int T^t_\phi \, dV \ , \tag{4}$$

where $dV$ is the volume element in the $t = const$ hypersurface. In addition, one can compute the ratio $m$ of the disk rest mass ($= \int \sigma\, u^t \, dV$) to the total mass $M$. We used this ratio as a global parameter characterizing self-gravity of the disk. The metric functions are defined in a finite grid in polar spheroidal coordinates outside the black hole horizon. In our finest grid, the horizon is identified with the inner boundary of the domain for the radial coordinate. The outer boundary should be at asymptotically flat infinity. This is not exactly possible in a numerical grid with the radial coordinate used here; therefore we chose the outer edge of integration domain to be at $r \equiv r_\star \approx 100\, M$ which is our "numerical infinity". In contrast to previous works (Lanza 1992b) we had to use significantly finer grid (typically $64 \times 128$ at the highest level of the multigrid scheme) because the solution of the geodesic equation requires to calculate derivatives of the metric functions numerically.

When the metric functions are used to integrate the light rays and to construct the line profile, one needs to trace the rays from the place of emission to infinity where the observer is located. Therefore an extension of the functions to infinity is necessary. This is done by approximating the solution by the Kerr metric for $r > r_\star$. The leading terms in the radial expansion of the metric functions of both the inner (numerical) and the outer (Kerr) spacetimes have identical forms and they should also take the same value when the same values of the total mass and total angular momentum are used in both regions. We identified only the first terms in the asymptotic expansion of the two metrics at $r_\star$. In other words, the numerical solution is applied inside the sphere $r = r_\star$ while the Kerr metric is assumed outside this sphere. The error of this approximation is of the order $\mathcal{O}(1/r^3)$ (Bardeen 1973a; Misner et al. 1973) and it does not affect the final spectral line profiles.

Our disks are confined completely within the region $r < r_\star$ with the inner edge close to the marginally stable orbit. In models of AGN, a fraction of spectral lines comes from regions farther away from the black hole where general relativistic effects are negligible. This part of the light can be included by applying calculations described in previous papers that ignore gravitational effects of the disk material.

### 2.2. The observed profile of a spectral line

The position of an observer at spatial infinity with respect to a source of radiation is characterized by the inclination angle $\theta_0$, that is the angle between the line of sight and the axis of symmetry of the system. The self-gravity of the disk influences not only the photon trajectory and redshift but also production of radiation from the disk itself. Within the approximation of the standard disk model, only total radiative flux from the disk plane is dominantly influenced because the standard model relates the total flux to the shear of the disk material at geodesic motion (see detailed discussion in Page & Thorne 1974; Shapiro & Teukolsky 1983). The spatial distribution of the radiative intensity (limb-darkening) and the frequency profile are likely to be less affected by the self-gravity of the disk because we consider here moderate matter densities only.

In order to transport photons from the disk to the observer, we solved the geodesic equation for light rays emitted from the disk that become deflected to $\theta_0$ asymptotically. The integration of the geodesic equation in the inner region requires appropriate interpolations of the metric functions in the $r\theta$-plane while in the outer region we have taken the Kerr metric and employed Carter's (1968) separation of the geodesic equation. The radiative intensity $I(\nu)$ in the observer frame is related to the intensity $I_{\rm d}(\nu_{\rm d})$ emitted by an element of the disk surface by the relation:

$$I(\nu) = g^3\, I_{\rm d}(\nu_{\rm d})\,. \tag{5}$$

Photons emitted with frequency $\nu_{\rm d}$ in the disk-corotating frame are detected with the redshifted or blueshifted frequency $\nu = g\nu_{\rm d}$. One can derive the explicit form of the $g$-factor:

$$g = \frac{e^{\tilde{\nu}}\sqrt{1-v^2}}{1-\Omega_{\rm d}\xi}\,, \tag{6}$$

where the linear velocity $v$ of the disk corotating element with respect to the zero-angular-momentum observers is given in terms of the metric functions and the angular velocity $\Omega_{\rm d}$:

$$v = (\Omega_{\rm d} - \omega)\, e^{\psi-\tilde{\nu}}\ .$$

The value of $\Omega_{\rm d}$, the Keplerian angular velocity of circular orbits, also follows from the numerical solution of Einstein's equations described in Section 2. In eq. (6), $\xi$ is the photon specific angular momentum (Chandrasekhar 1983). The expression is evaluated at the point where the ray is emitted from the disk and fully accounts for gravitational and Doppler shifts acting on photons. A careful analysis of the $g$-factor in a pure Kerr solution with a Keplerian disk has been originally given by Cunningham (1975).

As for the problem of production of radiation, one needs information on how it is emitted in the disk-corotating frame. A description of the local emissivity which is sufficiently realistic and at the same time consistent with the disk model poses a separate and difficult task in the astrophysiscs of accretion disks. We followed the scheme of the standard model in which the intensity of the emitted radiation at the disk surface is contructed indirectly from the total radiative flux $F(r)$ emitted by the disk at radius $r$ (Thorne 1974):

$$F(r) \propto e^{-(\tilde{\nu}+\psi+\mu)} \frac{\partial \Omega_{\rm d}}{\partial r} \left( E_{\rm d}^{\dagger} - \Omega_{\rm d} L_{\rm d}^{\dagger} \right)^{-2} \int_{r_{in}}^{r} \left( E_{\rm d}^{\dagger} - \Omega_{\rm d} L_{\rm d}^{\dagger} \right) \frac{\partial L_{\rm d}^{\dagger}}{\partial r} \, dr \, . \qquad (7)$$

$E_{\rm d}^{\dagger}$ and $L_{\rm d}^{\dagger}$ are the specific energy and the angular momentum of the equatorial circular geodesics measured at infinity, and $r_{in}$ is the inner radius of the disk. These quantities are provided by the black hole–disk self-consistent model described above. The radiative intensity is determined by

$$I_{\rm d}(\nu_{\rm d}; \vartheta_{\rm d}, r) = F(r) \, \varphi_1(\nu_{\rm d}) \, \varphi_2(\vartheta_{\rm d}) \, , \qquad (8)$$

where $\varphi_1(\nu_{\rm d})$ is the emissivity profile in frequency, $\varphi_2(\vartheta_{\rm d})$ is the limb-darkening law, and $\vartheta_{\rm d}$ is the angle between the ray and the local normal to the disk in the corotating frame. In the specific examples which are described below we considered

- isotropic distribution of the radiative intensity or a simple linear dependence on $\cos \vartheta_{\rm d}$ which is motivated by elementary results for the Compton scattering on the electrons (see, e.g., Thorne 1974).

- for the emitted profile in frequency we used Gaussian and Lorentzian profiles or a $\delta$-function line. The first two profiles have well-known physical motivations while the observed profile of a $\delta$-function line can serve as Green's function in calculating more complicated cases.

Locally generated radiation of the disk is then transported along photon rays to the observer. Finally, the observed flux in the interval $(\nu, \nu + d\nu)$ of the redshifted frequency is obtained by integrating $I(\nu)$ across the observer plane (Appendix).

## 3. RESULTS AND CONCLUSIONS

We computed a number of line profiles parametrized by the ratio $m$ of the disk rest mass to the total (black hole and disk) mass $M$, by the specific total angular momentum ($a \equiv J/M^2$), and by the inclination angle ($\theta_0$). The resulting profiles are modified by the light bending when the observer inclination is large. In addition, given the assumptions for constructing the disk model, radiation produced by the disk is different when the self-gravity of a (moderately massive) disk is properly accounted. We considered different regions where the line is formed and different emissivity laws. In order to understand the effects involved in forming the predicted line profiles we first considered a thin disk made by a test fluid around a Kerr black hole. Figure 1 shows the observed flux normalized to its maximum value as a function of the observed frequency normalized to frequency for which the local emissivity is maximum. In this case the self-gravity of the disk was ignored, analogously to the approximation used previously by other authors (Laor 1991; Kojima 1991; Stella & Campana 1991). We show these profiles for later comparison with the configurations in which the self-gravity of the disk is included. The cases (a)–(e) correspond to lines emitted isotropically with a Gaussian profile in the local rest frame of the disk material. The inclination of the observer increases from 20° for (a) to 85° for (e). Analogously, the cases (f)–(j) correspond to the lines with an anisotropic distribution of radiated intensity: $I_{\rm d}(\vartheta_{\rm d};\nu_{\rm d}) \propto 1 + 2.06\cos\vartheta_{\rm d}$ (Thorne 1974). We assumed that the emitting region is corotating with the disk. Besides the gravitational redshift of the line toward lower observed frequencies, the two evident peaks in the profiles are due to the Doppler effect. It is easy to understand that the peak at higher energy is more pronounced compared to the one at lower energy because the redshift factor ($g = \nu/\nu_{\rm d}$) enters with a third power into the formula for the observed flux [cf. eqs. (5) and (A1)]. Some authors report on the existence of additional extrema in predicted line profiles for high inclinations and interpret them as the effect of the light focusing by the black hole (Bao et al. 1994). We also did observe similar peaks in our profiles, however, this property depends also on the distribution of the emissivity along the disk. A phenomenological power law decrease of the emissivity as a function of radius, which is a usual assumption for intermediate and large distances from the black hole, favours additional peaks due to the light bending because the maximum of the disk emissivity coincides with the innermost disk region. These peaks are less apparent if the disk emissivity is small at the inner edge as it is for example in the standard disk model. For locally anisotropic radiation, which is beamed in the direction normal to the disk plane, the central peak becomes more pronounced when the inclination is $\gtrsim 80°$. The shape of the line profiles is nearly independent of the angular momentum parameter $a$ when the same local disk emissivity is used.

In Figure 3 the typical effects of the self-gravity of the disk on the line profile are shown. In this case $m = 0.01$. The self-gravity modifies the radial dependence of the function $\Omega_{\rm d}$ (Lanza 1992b) which determines the radiative flux generated by the disk in Page & Thorne's (1974) approximation. It is interesting to notice that the local emissivity has its maximum farther away from the black hole by several garvitational radii, compared to a test disk in the Kerr metric. This fact may be important for the discussion of the short-term variability in AGN. Other details of the disk emissivity are identical with those in Fig. 1. The effect of light bending can be distinguished by comparing profiles with no contribution from photons that are emitted from the opposite part of the disk with respect to the observer, i.e. neglecting the first image photons (Figure 3), an the profiles with the contribution from the first image photons included (Figure 5). The first case appears to be astrophysically more relevant because the outer edge of an accretion disk is presumably located at the distance $\lesssim 10^5$ gravitational radii from the centre and the disk is optically thick (however, cf. Bao et al. 1994 who suggest that the optical thickness of the disk material decreases significantly near the black hole so that the first image photons may become important). Only photons that cross the equatorial plane outside the disk and eventually reach the observer are considered in Fig. 5. Again, one can see that particularly with high inclinations the effect of the disk gravity is significant. Figure 7 shows how the equivalent width of the line is affected by increasing the mass of the disk. We plotted the relative difference in the equivalent width defined as

$$\delta\,\mathrm{EW}(m) \equiv \frac{\mathrm{EW}(m=0)-\mathrm{EW}(m)}{\mathrm{EW}(m=0)+\mathrm{EW}(m)}. \tag{9}$$

Two cases are shown for different inclination angles. For $m = 0.01$ the relative difference in the equivalent width is of the order of 1–3 per cent. Analogously to that, the relative centroid energy is decreased by 1 per cent. Typically, the Doppler broadening results in large equivalent widths which suggest that our results can be relevant for AGN with wide spectral features. Examples of Seyfert II galaxies with wide Fe lines have been recently reported (Awaki et al. 1994): NGC 1068 has a feature with EW = 1.0 keV, Mkn3 has EW = 0.86 keV, and NGC 4388 has EW = 0.642 keV.

Further details and a more extensive sample of predicted line profiles with different disks and different emissivity laws can be found in the electronic preprint library *astro-ph@babbage.sissa.it* or can be requested directly from the authors. We expect that a direct quantitative comparison with the observations will be possible in the future when data from the high resolution X-ray observatory are available. In particular, the XMM satellite is expected to reach the resolution of $E/\Delta E \approx 40$ at $E \approx (6-7)\,\mathrm{keV}$, and the order of magnitude better resolution at $E \approx 1\,\mathrm{keV}$.

We are grateful to A. Treves and R. Stark for reading the manuscript and for useful comments and suggestions. D.V. acknowledges SISSA, Trieste for the kind hospitality, and OCA/CERGA, Grasse where he stayed when completing this work thanks to the

H. Poincaré fellowship. Partial support by the Czech grant GA CR 205/94/0504 and by the Italian M.U.R.S.T. is acknowledged.

## APPENDIX

In this Appendix we present technical details concerning the numerical technique which was used to integrate null geodesics and to calculate the detected line profile. We denote $\mathbf{n}(\theta_0)$ the unit vector in the direction to a distant observer with inclination angle $\theta_0$ and we introduce the polar coordinates $\rho$, $\varphi$ in the plane $\mathcal{P}$ perpendicular to $\mathbf{n}(\theta_0)$ at spatial infinity. The origin $\rho = 0$ corresponds to the principal null ray along the $\theta_0$-direction. The radiative flux measured by the observer at frequency $\nu$ can be evaluated in the form

$$F(\nu;\theta_0) = \int_{\mathcal{P}} I(\rho,\varphi;\nu)\,\rho\,d\rho\,d\varphi\,. \tag{A1}$$

Here $I(\rho,\varphi;\nu)$ stands for the radiative intensity carried by the photon which intersects $\mathcal{P}$ at the point $(\rho,\varphi)$. The observed radiative intensity $I$ and the detected frequency $\nu$ are related to corresponding values $I_{\rm d}$ at $\nu_{\rm d}$ at the point of emission by eq. (5).

The photon ray is specified by the associated specific angular momentum $\xi$ and Carter's constant $\eta$ which enter directly into the equation of null geodesic in its standard form (Chandrasekhar 1983). They are related to coordinates $\rho$, $\varphi$ of the point where the ray intersects $\mathcal{P}$ by

$$\xi = \rho\sin\theta_0\sin\varphi\,, \qquad \eta = \rho^2\cos^2\varphi - a^2\cos^2\theta_0 + \xi^2\cot^2\theta_0\,. \tag{A2}$$

The relevant geodesic equation takes the form

$$\int_{\infty}^{r_\star/M} \frac{dr}{R(r/M;\xi,\eta,a)^{1/2}} = \int_{\mu_0}^{\mu_\star} \frac{d\mu}{\Theta(\mu;\xi,\eta,a)^{1/2}}\,, \tag{A3}$$

where $\mu_0 = \cos\theta_0$ and $\mu_\star = \cos\theta_\star$, $\theta_\star = \theta(r_\star)$,

$$\begin{aligned} R(r;\xi,\eta,a) &= r^4 + \left(a^2 - \xi^2 - \eta\right) r^2 + 2\left[\eta + (\xi - a)^2\right] r - a^2\eta\,, \\ \Theta(\mu;\xi,\eta,a) &= \eta - \left(\xi^2 + \eta - a^2\right)\mu^2 - a^2\mu^2\,. \end{aligned}$$

Eq. (A3) can be solved for $\theta_\star$ in terms of Jacobian elliptic functions. We used tables of elliptic integrals by Byrd & Friedman (1971; cf. a correction by Viergutz 1993) and developed an efficient code to treat the problem. Similar code for integrating time-like geodesics was described in detail by Karas & Vokrouhlický (1994).

To summarize this part: specifying the ray by $\rho$ and $\varphi$, we find $\theta_\star$ of its intersection with $r = r_\star$. This is the initial position from which the numerical integration inside $r < r_\star$ starts. It is instructive to demonstrate an error that one would make by ignoring the light bending outside $r_\star$: Figure 8a shows the difference between $\theta_\star$ which is obtained through the above described procedure and the corresponding value in the flat spacetime, while Figure 8b shows the difference in $\theta_\star$ between $a = 0.998$ and $a = 0$ cases. It is evident that the error is much greater in the first case, in agreement with the fact that line profiles may be significantly affected by bending of light but the difference between rotating and non-rotating cases is only marginal. To illustrate the effect clearly, we adopted high value of the Kerr angular momentum parameter, $a = 0.998$, $r_\star = 50\ M$ for the inner-to-outer region boundary, and the inclination angle 80°.

For $r < r_\star$, we considered geodesic equation in the form

$$\dot{p^r} + \Gamma^r_{\alpha\beta}\,(\tilde{\nu}, \psi, \mu; r, \theta)\, p^\alpha p^\beta = 0\ , \qquad p^r = \dot{r}\ , \tag{A4}$$

$$\dot{p^\theta} + \Gamma^\theta_{\alpha\beta}\,(\tilde{\nu}, \psi, \mu; r, \theta)\, p^\alpha p^\beta = 0\ , \qquad p^\theta = \dot{\theta}\ , \tag{A5}$$

where Christoffel's symbols $\Gamma$ correspond to the general form of axially symmetric metric (1) with indeces $\alpha$, $\beta$ spanning $0, \ldots, 3$. This sytem of equations can be solved in a straightforward way because there are two conserved quantities, $p_t$ and $p_\phi$, existence of which follows from the stationarity and axial symmetry of the spacetime (1). All quantities are calculated in a finite grid of $r$, $\theta$ and interpolated by bivariate six-point formula (Abramowitz & Stegun 1972).

---

FIGURE CAPTIONS

Figure 1—Observed flux normalized to maximum versus observed frequency normalized to the frequency for which the local emissivity is maximum. The observed line profiles are for a test disk around a Kerr black hole; $a = 0.998$, $r_{\rm in} = 5.5\ M$ (marginally stable orbit), $r_{\rm out} = 85\ M$. The left column corresponds to the line which radiates isotropically with the Gaussian profile in the local disk-corotating frame (the line width 0.6 in dimensionless units). The right column corresponds to an anisotropic emission with the same local profile and width. The observer inclination $\theta_0$ : (a,f) 20°; (b,g) 40°; (c,h) 60°; (d,i) 80°; (e,j) 85°. Solid line corresponds to the contribution of the whole disk. The contributions from four equally wide regions between $r_{\rm in}$ and $r_{\rm out}$ in the disk are shown by different line styles; with increasing $r$: dashes; dots and long dashes; dots and short dashes; dots.

Figure 2—As in Fig. 1 but for a Lorentzian profile of the line in the disk corotating frame.

Figure 3—As in Fig. 1 but for a self-gravitating disk; $m = 0.01$, $a = 0.5$, $r_{\rm in} = 8\ M$, $r_{\rm out} = 20\ M$. Gaussian profile of the line in the disk corotating frame. For further description see Section 3.

Figure 4—As in Fig. 3 but for a Lorentzian profile of the line in the disk corotating frame.

Figure 5—As in Fig. 3 but with the contribution from rays bended across the equatorial plane (the first order image) included. Gaussian profile of the line in the disk corotating frame.

Figure 6—As in Fig. 5 but for a Lorentzian profile of the line in the disk corotating frame.

Figure 7—Relative difference $\delta$ EW of the equivalent width of the line as a function of the ratio $m$ of the disk mass to the total mass of the system. The inclination angle is 20° (solid line) and 85° (dashed line).

Figure 8—This figure shows the absolute magnitude of the difference $\delta\theta$ at $r = r_\star$ as a function of $\varphi$ for different values of $\rho$ ($\rho$ goes linearly from 0 to $r_\star$; the curve with maximum $|\delta\theta|$ corresponds to $\rho \longrightarrow r_\star$). The case (a) shows $|\delta\theta|$ between geodesics in the Kerr metric and straight lines ignoring the effects of the light bending; the case (b) compares geodesics in the Kerr metric to those in the Schwarzschild metric. The most pronounced errors which occur in (a) for large $\rho$ are eliminated in (b) (notice a different scale on the vertical axis in both Figures). This is due to the fact that the shape of photon trajectories in the Schwarzschild metric approximates the correct shape adequately for $r \gtrsim r_\star$. In practice, angular momentum of the black hole is important in studying the disk structure but otherwise it has a little direct effect on observed profiles of spectral lines.

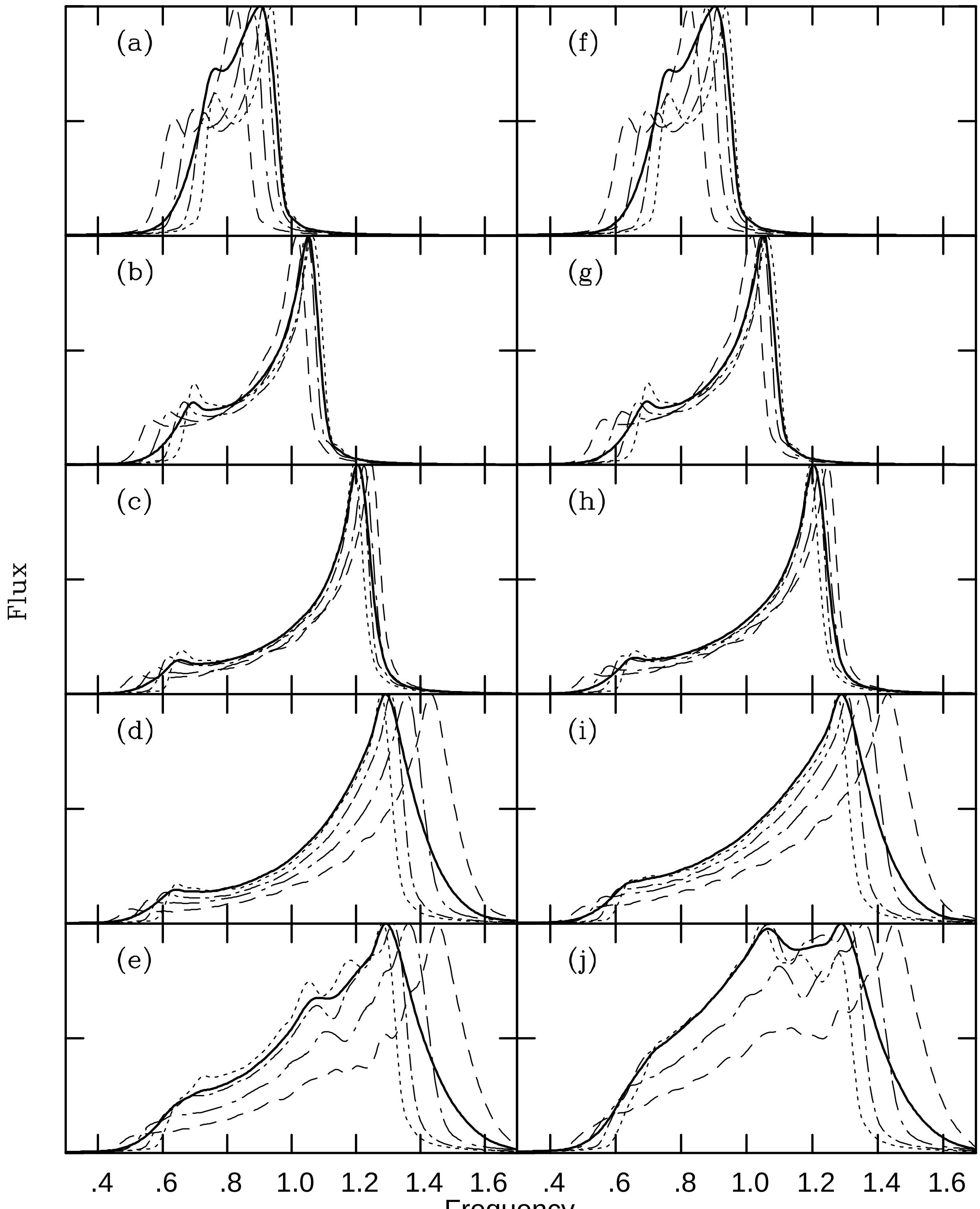

Fig. 1.—

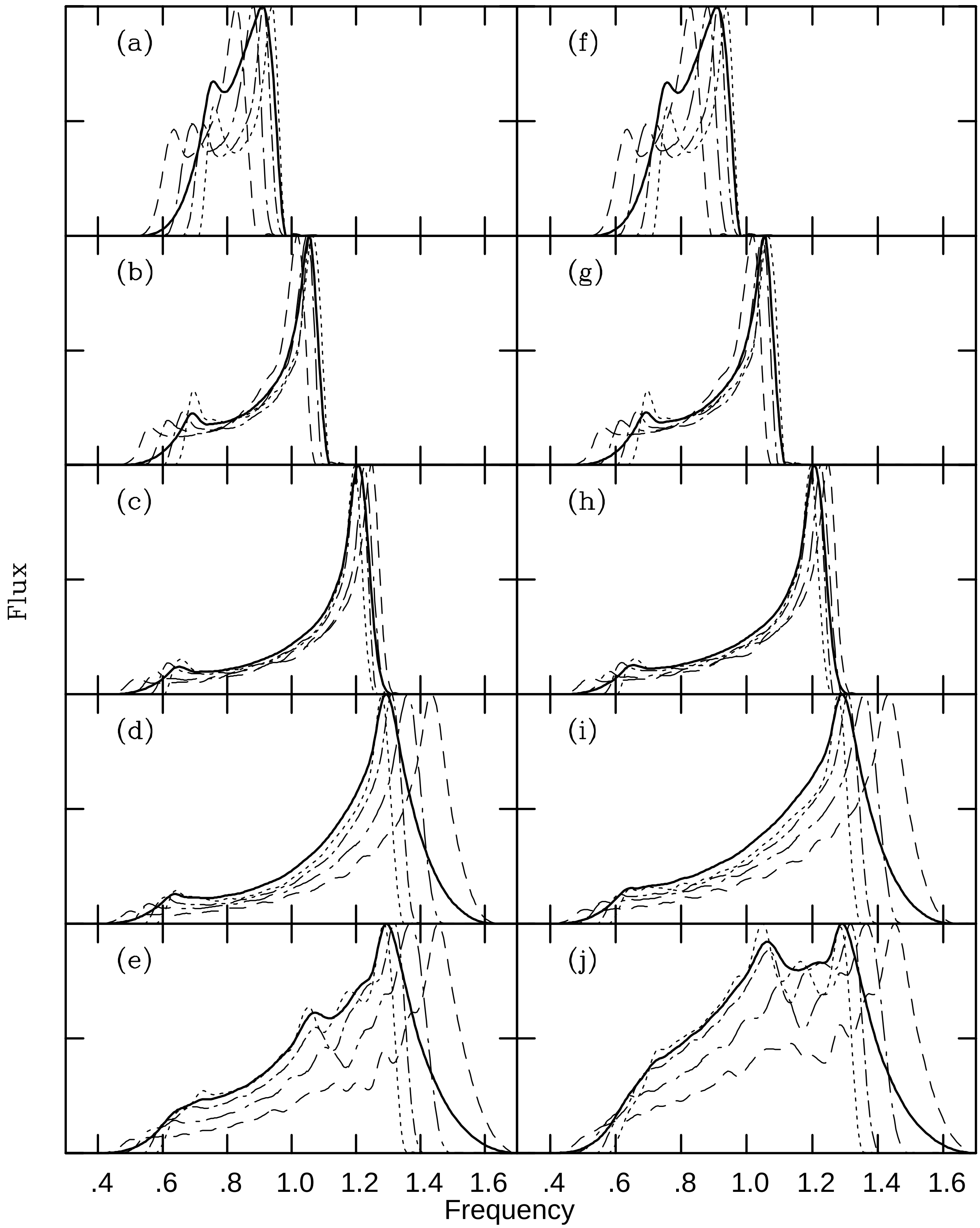

Fig. 2.—

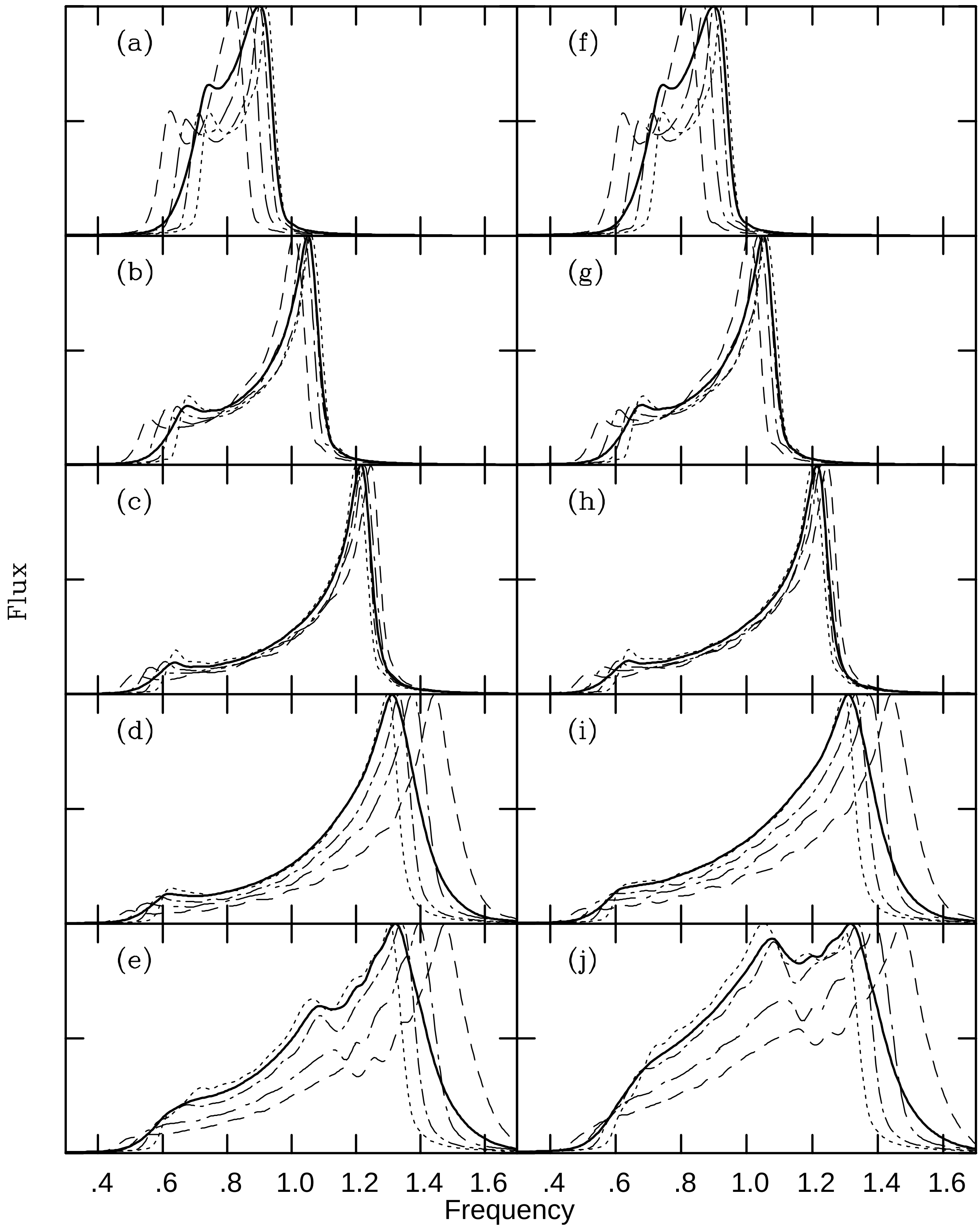

Fig. 3.—

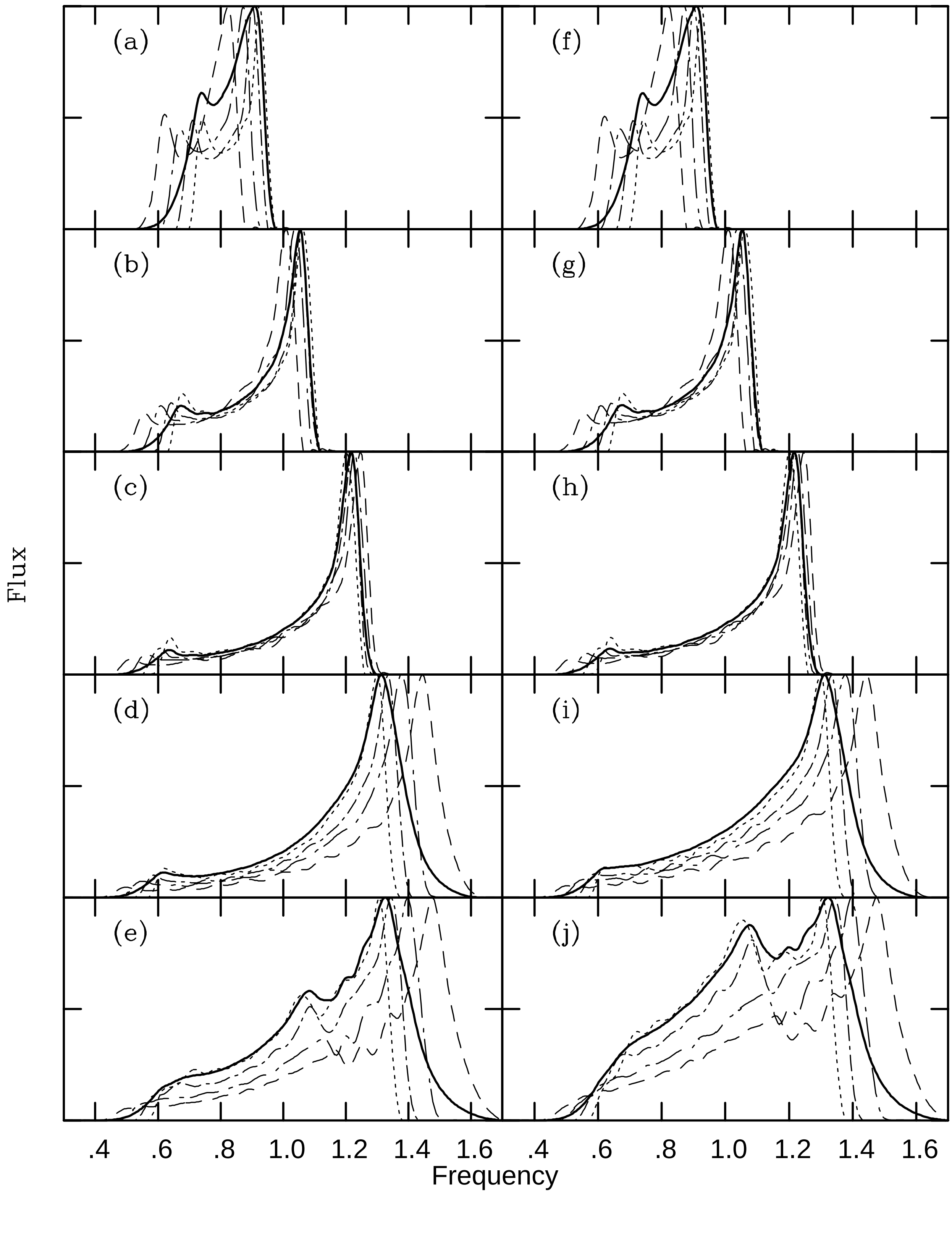

Fig. 4.—

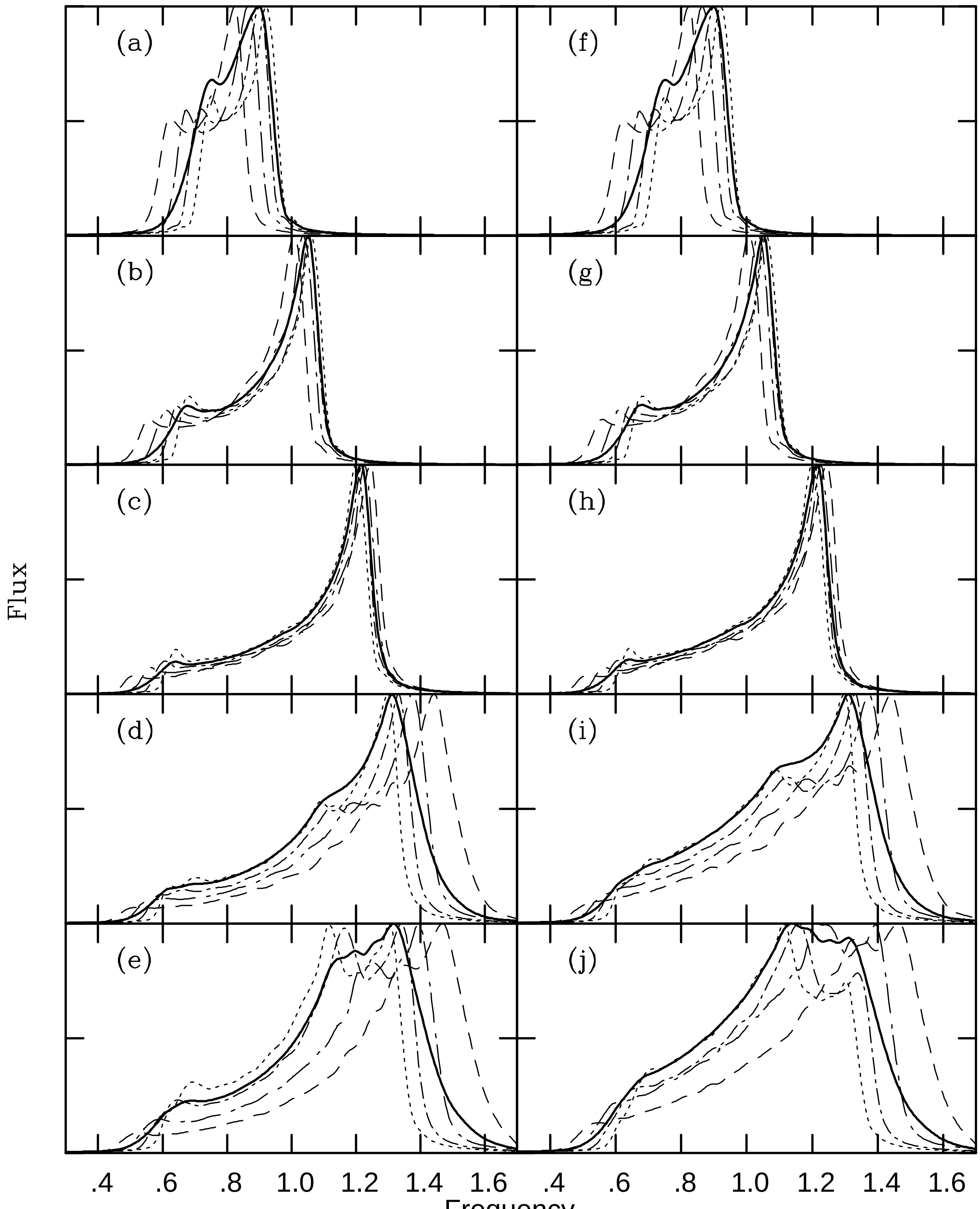

Fig. 5.—

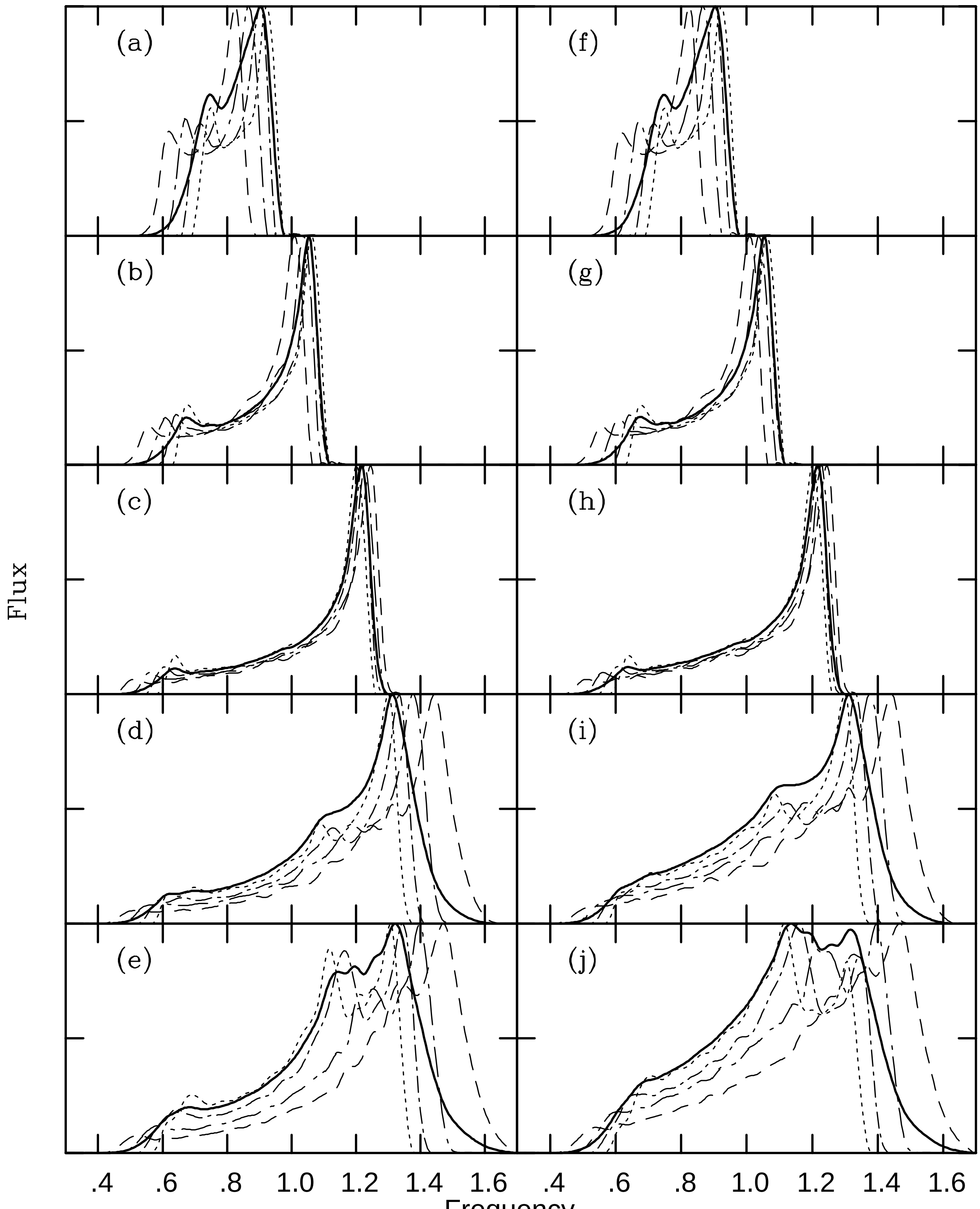

Fig. 6.—

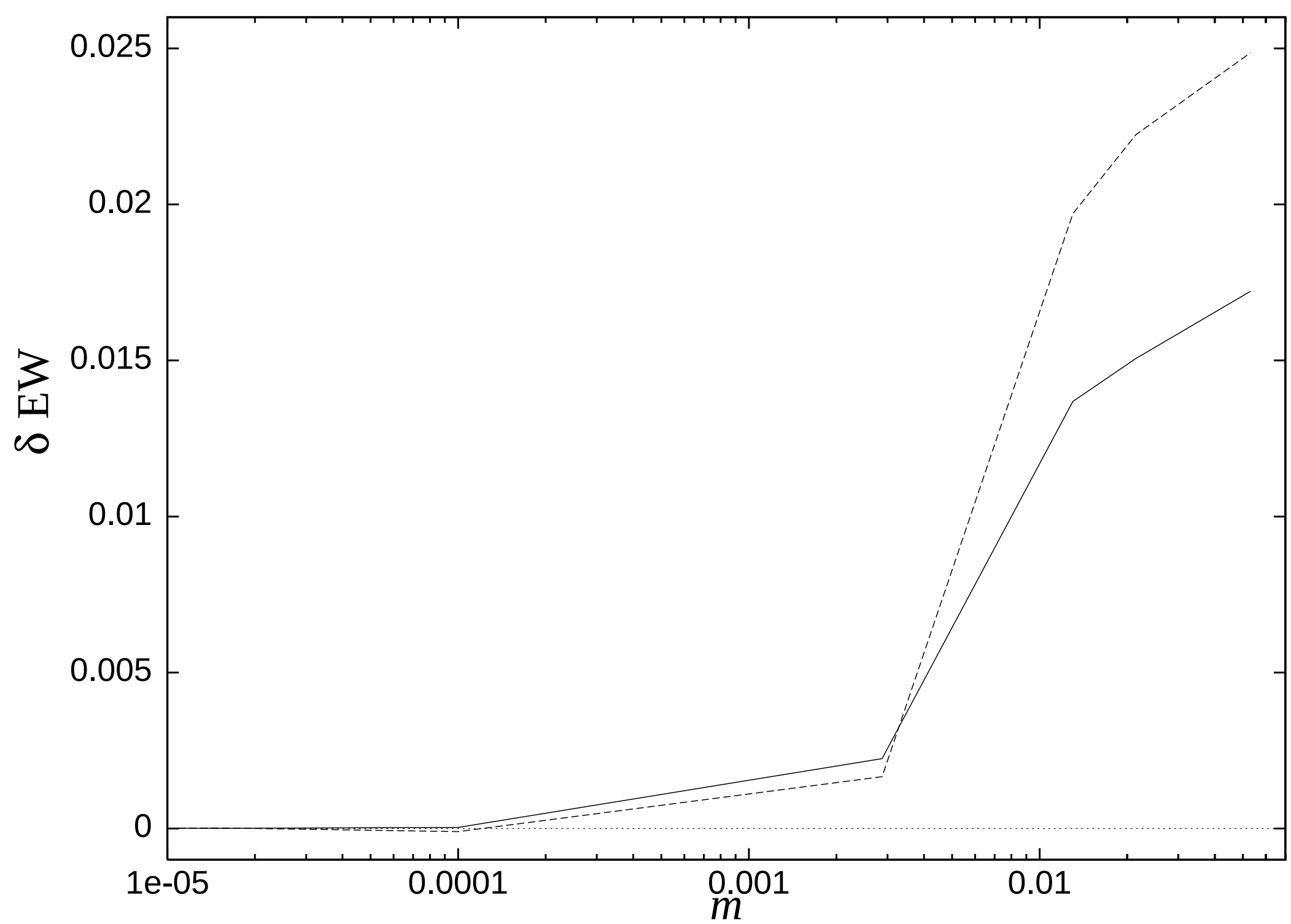

Fig. 7.—

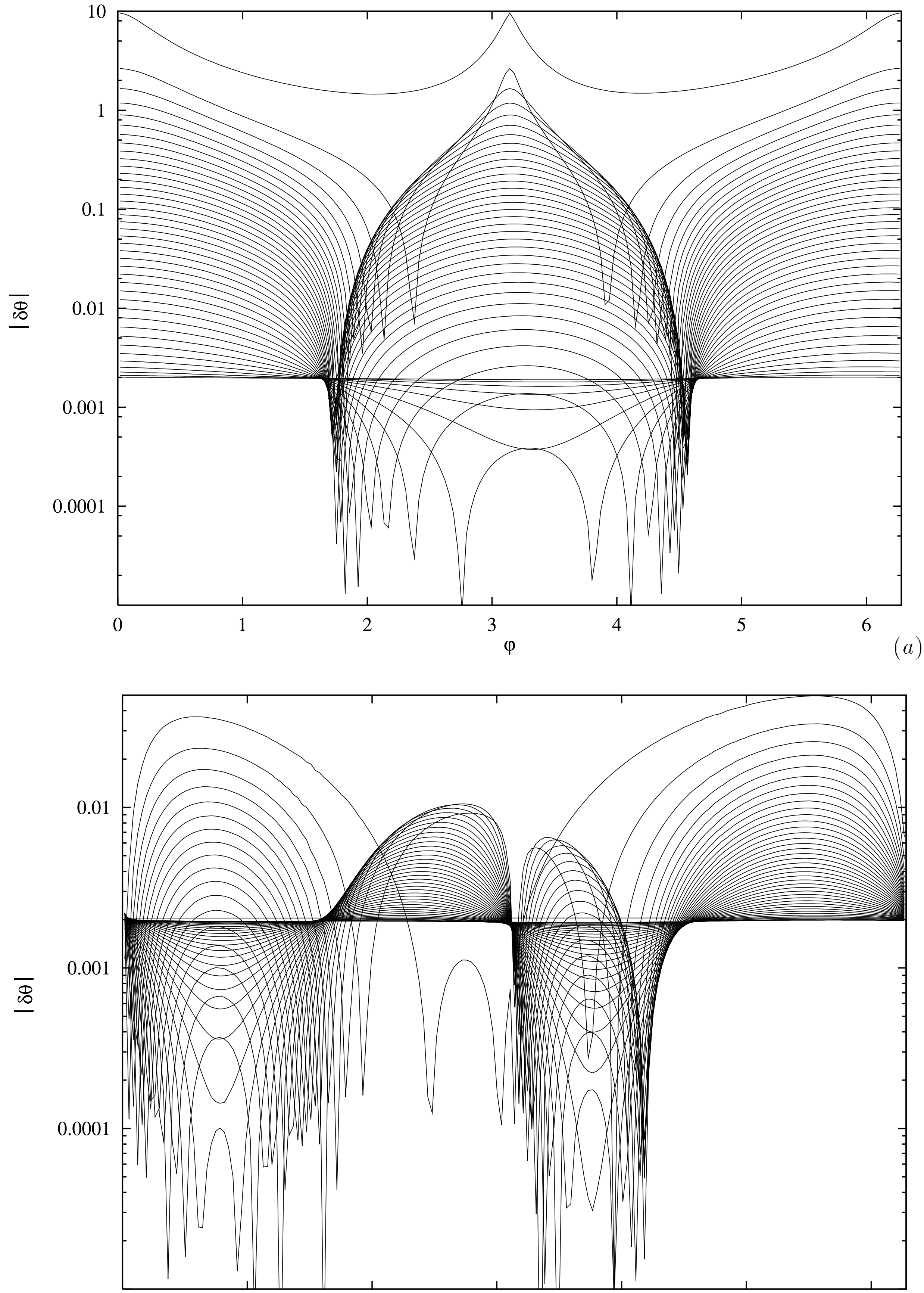
10
1
0.1
0.01
0.001
0.0001
|δθ|
0
1
2
3
4
5
6
φ
(a)
0.01
0.001
0.0001
|δθ|
0
1
2
3
4
5
6
φ
(b)

Fig. 8.—